\documentclass[pdftex,twocolumn,epjc3]{svjour3}   
\RequirePackage[T1]{fontenc}

\smartqed  % flush right qed marks, e.g. at end of proof

\RequirePackage{graphicx}
\RequirePackage{epstopdf}
\RequirePackage{mathptmx}      % use Times fonts if available on your TeX system
\RequirePackage{flushend}
\RequirePackage[numbers,sort&compress]{natbib}
\RequirePackage[colorlinks,citecolor=blue,urlcolor=blue,linkcolor=blue]{hyperref}
\usepackage{amsmath}
\usepackage{amssymb}
\journalname{Eur. Phys. J. C}

\begin{document}
\title{Rest frame vacuum of the Dirac field on spatially flat FLRW spacetimes}

\author{Ion I. Cot\u aescu %\inst{1}
\thanks{e-mail: i.cotaescu@e-uvt.ro}}
\institute{West University of Timi\c soara, V. P\^ arvan Ave. 4, RO-300223, Timi\c soara, Romania}
\date{Received: date / Revised version: date}

\maketitle

\begin{abstract}
In quantum theory of the free Dirac field on spatially flat FLRW spacetimes we introduce a new type of vacuum able to separate the positive and negative frequencies in the rest frames. This is refereed as the rest frame vacuum shoving that this differs from the adiabatic vacuum apart from the Minkowski spacetime where these two vacua coincide.    

Pacs: 04.62.+v
\end{abstract}

\section{Introduction}

In general relativity, the problem of defining the vacuum  is related to the separation of the positive and negative frequencies of the free quantum fields in the absence of a conserved energy operator commuting with the field equation. Thus many vacua can be considered simultaneously in the non-perturbative approaches as, for example, the cosmological creation of elementary particles of different spins \cite{P1,P2,S1}. In this approach many studies were devoted to the scalar fields on various manifolds, discussed in a vaste literature summarized in Ref. \cite{BD}, leading to the common accepted definition of the adiabatic vacuum on the FLRW manifolds. 

The Dirac field was studied maily  in the local charts with spherical coordinates of the FLRW spacetimes where its time evolution is governed by a pair of time modulation functions that, in general, cannot be solved analytically since they satisfy equations of oscillators with variable frequencies  \cite{c1,c2,c3,c4,c5}. Nevertheless, in the de Sitter case we succeeded to solve for the first time the Dirac equation in momentum-helicity \cite{CD1,CD2} or momentum-spin \cite{CD3} representations,  separating  the frequencies according to the general prescription of the adiabatic vacuum. These solutions are expressed in terms of normalized fundamental spinors complying with the charge conjugation that can be used for calculating physical effects in our de Sitter QED in Coulomb gauge \cite{CQED,Cr1,Cr2}.

The momentum-spin representation offers us the opportunity of analyzing the fundamental spinors in the rest frame where the momentum vanishes. We observed that in this limit the fundamental spinors defined according to the adiabatic vacuum do not coincide with  the normalized spinors calculated directly in the rest frames \cite{CD3}. This suggests us that a new vacuum must be introduced in order to assure a complete separation between particles and antiparticles in the rest frames. This paper is devoted to this vacuum, called the rest frame vacuum, that can be defined for the Dirac field minimally coupled to the gravity of any $(3+1)$-dimensional spatially flat FLRW spacetime thanks to our method of analyzing the fundamental spinors \cite{CIRdS,CIRF}.

We devote the next two sections to  the free Dirac field on the spatially flat  FLRW manifolds pointing out that its fundamental spinors depend on four time modulation functions which satisfy the normalization and charge conjugation conditions. In the next section we generalize the definition of the well-known adiabatic vacuum and we introduce the new rest frame vacuum. In the fifth section we present three examples, the Minkowski spacetime where the adiabatic and rest frame vacua coincide, and two backgrounds where these vacua are different, namely the de Sitter spacetime and a new spatially flat FLRW manifold with a Milne type scale factor we studied recently \cite{CIRF,prep}. Finally we present our concluding remarks.
 
\section{Dirac's field on FLRW spacetimes}

Let us consider the $(1+3)$-dimensional spatially flat FLRW manifolds, $M$, where we choose the comoving local FLRW charts $\{t,{\bf x}\}$ whose coordinates $x^{\mu}$ (labeled  by the natural indices $\mu,\nu,...=0,1,2,3 $) are formed by the proper (or cosmic) time, $t$, and Cartesian space coordinates, $x^i$  ($i,j,k...=1,2,3$),  for which we may use the vector notation ${\bf x}=(x^1,x^2,x^3)$.  The FLRW geometry is given by a smooth scale factor $a(t)$ defining the conformal time as,
\begin{equation}
t_c=\int \frac{dt}{a(t)}~\to~  a(t_c)=a[t(t_c)]\,.
\end{equation}
and determining the line elements,  
\begin{eqnarray}
ds^2=g_{\mu\nu}(x)dx^{\mu}dx^{\nu}&=&dt^2-a(t)^2 d{\vec x}\cdot d{\vec x}\nonumber\\
&=&a(t_c)^2(dt_c^2-d{\vec x}\cdot d{\vec x})\,,
\end{eqnarray}
of the FLRW chart and of the  associated conformal flat charts, $\{t_c,{\bf x}\}$. 

In these charts, we chose the diagonal tetrad gauge in which the vector fields $e_{\hat\alpha}=e_{\hat\alpha}^{\mu}\partial_{\mu}$ defining the local orthogonal frames,  and the 1-forms $\omega^{\hat\alpha}=\hat e_{\mu}^{\hat\alpha}dx^{\mu}$ of the dual coframes (labeled by the local indices, $\hat\mu,\hat\nu,...=0,1,2,3$) are defined as 
\begin{eqnarray}
&e_0=\partial_t=\frac{1}{a(t_c)}\,\partial_{t_c}\,,\qquad & \omega^0=dt=a(t_c)dt_c\,,\label{tetrad} \\
&~~~e_i=\frac{1}{a(t)}\,\partial_i=\frac{1}{a(t_c)}\,\partial_i\,, \qquad & \omega^i=a(t)dx^i=a(t_c)dx^i\,,
\end{eqnarray}
in order to preserve the global $SO(3)$ symmetry  allowing us to use systematically the  $SO(3)$ vectors. We remind the reader that the metric tensor of $M$ can be expressed now as $g_{\mu\nu}=\eta_{\hat\alpha\hat\beta}\hat e^{\hat\alpha}_{\mu}\hat e^{\hat\beta}_{\nu}$ where $\eta={\rm diag}(1,-1,-1,-1)$ is the Minkowski metric. 

In this tetrad-gauge, the massive Dirac field $\psi$ of mass $m$  satisfies the field equations $(D_x-m) \psi (x)=0$ given by the Dirac operator 
\begin{equation}\label{ED}
D_x=i\gamma^0\partial_{t}+i\frac{1}{a(t)}\gamma^i\partial_i
+\frac{3i}{2}\frac{\dot{a}(t)}{a(t)}\gamma^{0}\,.
\end{equation}
This operator is expressed in terms of the Dirac $\gamma$-matrices  and the scale factor $a(t)$ and its derivative denoted as $\dot{a}(t)=\partial_ta(t)$. It is known that the terms of these operators depending on the Hubble function $\frac{\dot{a}}{a}$ can be removed at any time by substituting $\psi \to [a(t)]^{-\frac{3}{2}}\psi$. Similar results can be written in the conformal chart.

The general solution of the Dirac equation  may be written as a mode integral, 
\begin{eqnarray}
\psi(t,{\bf x}\,)& =& 
\psi^{(+)}(t,{\bf x}\,)+\psi^{(-)}(t,{\bf x}\,)\nonumber\\
& =& \int d^{3}p
\sum_{\sigma}[U_{{\bf p},\sigma}(x){\frak a}({\bf p},\sigma)
+V_{{\bf p},\sigma}(x){\frak b}^{\dagger}({\bf p},\sigma)]\,,\label{p3}
\end{eqnarray}
in terms of the fundamental spinors $U_{{\bf p},\sigma}$  and  $V_{{\bf p},\sigma}$ of positive and respectively negative frequencies which are plane waves solutions of the Dirac equation depending on the conserved momentum ${\bf p}$ and an arbitrary polarization $\sigma$. These spinors  form an orthonormal  basis being related  through the charge conjugation, 
\begin{equation}\label{chc}
V_{{\bf p},\sigma}(t,{\bf x})=U^c_{{\bf p},\sigma}(t,{\bf x}) =C\left[{U}_{{\bf p},\sigma}(t,{\bf x})\right]^* \,, \quad C=i\gamma^2\,,
\end{equation}
(see the  Appendix A), and satisfying the orthogonality relations
\begin{eqnarray}
\langle U_{{\bf p},\sigma}, U_{{{\bf p}\,}',\sigma'}\rangle &=&
\langle V_{{\bf p},\sigma}, V_{{{\bf p}\,}',\sigma'}\rangle=
\delta_{\sigma\sigma^{\prime}}\delta^{3}({\bf p}-{\bf p}\,^{\prime})\label{ortU}\\
\langle U_{{\bf p},\sigma}, V_{{{\bf p}\,}',\sigma'}\rangle &=&
\langle V_{{\bf p},\sigma}, U_{{{\bf p}\,}',\sigma'}\rangle =0\,, \label{ortV}
\end{eqnarray}
with respect to the relativistic scalar product \cite{CD1}
\begin{eqnarray}
\langle \psi, \psi'\rangle&=&\int d^{3}x
\sqrt{|g|}\,e^0_0\,\bar{\psi}(x)\gamma^{0}\psi(x) \nonumber\\
&=&\int d^{3}x\,
a(t)^{3}\bar{\psi}(x)\gamma^{0}\psi(x)\,, 
\end{eqnarray}
where $g={\rm det}(g_{\mu\nu})$ and $\bar{\psi}=\psi^+\gamma^0$ is the Dirac adjoint of $\psi$. Moreover,  this basis is supposed to be complete accomplishing the completeness condition  \cite{CD1}
\begin{eqnarray}
&&\int d^{3}p
\sum_{\sigma}\left[U_{{\bf p},\,\sigma}(t,{\bf x}\,)U^{+}_{{\bf p},\sigma}(t,{\bf x}\,^{\prime}\,)+V_{{\bf p},\sigma}(t,{\bf x}\,)V^{+}_{{\bf p},\sigma}(t,{\bf x}\,^{\prime}\,)\right] \nonumber\\
&&\hspace*{18mm}=a(t)^{-3}\delta^{3}({\bf x}-{\bf x}\,^{\prime})\,.\label{complet}
\end{eqnarray}
We obtain thus the orthonormal basis of the momentum representation  in which   the particle $({\frak a},{\frak a}^{\dagger})$ and antiparticle (${\frak b},{\frak b}^{\dagger})$ operators  satisfy the canonical anti-commutation relations \cite{CD1}.

\section{Fundamental spinors}

In the standard representation of the Dirac matrices (with diagonal $\gamma^0$) the general form of the fundamental spinors in momentum representation,   
\begin{eqnarray}
U_{\vec{p},\sigma}(t,\vec{x}\,)&=&\frac{e^{i\vec{p}\cdot\vec{x}}}{[2\pi a(t)]^{\frac{3}{2}}}\left(
\begin{array}{c}
u^+_p(t) \,
\xi_{\sigma}\\
u^-_p(t) \,
 \frac{{p}^i{\sigma}_i}{p}\,\xi_{\sigma}
\end{array}\right)
\label{Ups}\\
V_{\vec{p},\sigma}(t,\vec{x}\,)&=&\frac{e^{-i\vec{p}\cdot\vec{x}}}{[2\pi a(t)]^{\frac{3}{2}}} \left(
\begin{array}{c}
v^+_p(t)\,
\frac{{p}^i{\sigma}_i}{p}\,\eta_{\sigma}\\
v^-_p(t) \,\eta_{\sigma}
\end{array}\right)
\,,\label{Vps}
\end{eqnarray}
is determined by the modulation functions $u^{\pm}_p(t)$ and $v^{\pm}_p(t)$  that depend only on $t$ and  $p=|\vec{p}|$.   The Pauli spinors $\xi_{\sigma}$ and $\eta_{\sigma}= i\sigma_2 (\xi_{\sigma})^{*}$ must be correctly normalized,  $\xi^+_{\sigma}\xi_{\sigma'}=\eta^+_{\sigma}\eta_{\sigma'}=\delta_{\sigma\sigma'}$,  satisfying the completeness condition 
\begin{equation}\label{Pcom}
\sum_{\sigma}\xi_{\sigma}\xi_{\sigma}^+=\sum_{\sigma}\eta_{\sigma}\eta_{\sigma}^+={\bf 1}_{2\times 2}\,.
\end{equation}
In Ref. \cite{CD1} we considered the Pauli spinors of the momentum-helicity basis whose direction of  the spin projection is just that of the momentum $\vec{p}$. However, we can project the spin on an arbitrary direction, independent on $\vec{p}$, as in the case of the
{\em spin} basis  where the spin is projected on the third axis of the rest frame such that  $\xi_{\frac{1}{2}}=(1,0)^T$ and $\xi_{-\frac{1}{2}}=(0,1)^T$ for particles and $\eta_{\frac{1}{2}}=(0,-1)^T$ and $\eta_{-\frac{1}{2}}=(1,0)^T$ for
antiparticles \cite{CD3}. In what follows we work exclusively in this basis called  the momentum-spin basis. 

The modulation functions $u_p^{\pm}(t)$ and $v_p^{\pm}(t)$  can be derived   by substituting Eqs. (\ref{Ups}) and (\ref{Vps}) in the Dirac equation.  Then, after a few manipulation, we find the systems of the first order differential equations
\begin{eqnarray}
a(t)\left(i\partial_t\mp m\right)u_p^{\pm}(t)&=&{p}\,u_p^{\mp}(t)\,,\label{sy1}\\
a(t)\left(i\partial_t \mp m\right)v_p^{\pm}(t)&=&-{p}\,v_p^{\mp}(t)\,,\label{sy2}
\end{eqnarray}
in the chart with the proper time or the equivalent system in the conformal chart,
\begin{eqnarray}
\left[i\partial_{t_c}\mp m\, a(t_c)\right]u_p^{\pm}(t_c)&=&{p}\,u_p^{\mp}(t_c)\,,\label{sy1c}\\
\left[i\partial_{t_c} \mp m\, a(t_c)\right]v_p^{\pm}(t_c)&=&-{p}\,v_p^{\mp}(t_c)\,,\label{sy2c}
\end{eqnarray}
which govern the time modulation of the free Dirac field on any spatially flat FLRW manifold. Note that these equations are similar to those of the modulation functions of the spherical modes \cite{c1,c2,c3,c4,c5} but depending on different integration constants.

The solutions of these systems depend on  integration constants that must be selected according to the charge conjugation (\ref{chc}) which gives the mandatory condition
\begin{equation}\label{VU}
v_p^{\pm}(t)=\left[u_p^{\mp}(t)\right]^*\,.
\end{equation}
The remaining normalization constants can be fixed since the prime integrals of the systems (\ref{sy1}) and (\ref{sy2}), 
$\partial_t (|u_p^+|^2+|u_p^-|^2)=\partial_t (|v_p^+|^2+|v_p^-|^2)=0$, 
allow us to impose the normalization conditions
\begin{equation}
|u_p^+|^2+|u_p^-|^2=|v_p^+|^2+|v_p^-|^2 =1 \label{uuvv}\\
\end{equation}
which guarantee that Eqs.  (\ref{ortU}) and (\ref{ortV}) are accomplished. 

A special case is that of the rest frame where  the Dirac equation in momentum-spin representation for  ${\bf p}=0$ can be solved analytically carrying out the normalized fundamental spinors of the rest frame,
\begin{eqnarray}
U_{0,\sigma}(t,{\bf x})&=&\frac{e^{-i mt}}{[2\pi a(t)]^{\frac{3}{2}}}\left(
\begin{array}{c}
\xi_{\sigma}\\
0
\end{array}\right)\,,\label{Ur}\\
V_{0,\sigma}(t,{\bf x})&=&\frac{e^{i mt}}{[2\pi a(t)]^{\frac{3}{2}}}\left(
\begin{array}{c}
0\\
\eta_{\sigma}
\end{array}\right)\,,\label{Vr}
\end{eqnarray} 
which depend on the rest energy $E_0=m$.  Note that in the momentum-helicity representation the rest frame spinors cannot be defined since the helicity is related to a non-vanishing momentum.

\section{Adiabatic and rest frame vacua}

The modulation  functions, $u_p^{\pm}(t)$ or $u_p^{\pm}(t_c)$,  can be found by integrating  the systems (\ref{sy1}) or (\ref{sy1c})  in each particular case separately and imposing the charge conjugation and normalization conditions, (\ref{VU}) and respectively  (\ref{uuvv}).  However, these conditions are not enough for determining completely these  functions such that a supplemental physical hypothesis is required. This is just the criterion of separating the positive and negative frequencies defining thus the vacuum.    

The vacuum usually considered in Dirac theories is the traditional adiabatic vacuum intensively studied in the case of the scalar fields \cite{BD}. This can be defined for any FLRW manifold for which the scale factor satisfies the condition 
\begin{equation}\label{asc}
\lim_{t_c\to-\infty}a(t_c)=0\,.
\end{equation}
Then the asymptotic form of the system (\ref{sy1c}), 
\begin{equation}
i\partial_{t_c}u_p^{\pm}(t_c)=pu_p^{\mp}(t_c)\,,
\end{equation}
gives the behavior of the modulation functions, 
\begin{equation}
u_p^{\pm}(t_c)\sim c_1 e^{-ipt_c}\pm c_2 e^{ipt_c}\,,
\end{equation}
for $t_c\to -\infty$. According to the common definition,  the adiabatic vacuum is set when $c_2=0$  since then the modulation functions, $u_p^{+}(t_c)=u_p^{-}(t_c)$, describe a massless particle assumed to be of genuine positive frequency. Thus the general condition of selecting the adiabatic vacuum  of the Dirac field on FLRW spacetimes takes the simple form
\begin{equation}\label{adi}
u_p^{-}(t_c, m)=u_p^{+}(t_c,-m)
\end{equation}
and similarly for the functions $v_p^{\pm}(t_c)$.

The major difficulty of the adiabatic vacuum as defined above is that in the momentum-spin representation we cannot reach the rest frame limit.  Indeed, for $p\to 0$ the condition (\ref{adi}) gives the normalized functions
\begin{equation}
\lim_{p\to 0} u_p^{+}(t_c)=\frac{1}{\sqrt{2}}\,e^{-imt}\,, \quad \lim_{p\to 0}  u_p^{-}(t)=\frac{1}{\sqrt{2}}\,e^{imt}\,,
\end{equation}  
while the limit of $ \frac{{p}^i{\sigma}_i}{p}$ for $p\to 0$  is undetermined. Moreover, if we force this limit to zero we obtain a function $u^+_p$ with a different normalization factor \cite{CD3}. Therefore, the corresponding limits of the fundamental spinors will  differ from the correct rest spinors (\ref{Ur}) and (\ref{Vr}) mixing positive and negative frequencies. In other words, if we adopt the adiabatic vacuum then we find in the rest frames a  particle-antiparticle mixing which means that the separation of  the positive and negative frequencies in the asymptotic zone does not guarantee that this separation holds in any frame. 

The solution is to define a new vacuum able to separate the frequencies in any rest frame of the momentum-spin representation imposing the conditions
\begin{eqnarray}
\lim_{\vec{p}\to 0} U_{\vec p,\sigma}(t,{\bf x})&=&U_{0,\sigma}(t,{\bf x})\,,\label{Urf}\\
\lim_{\vec{p}\to 0} V_{\vec p,\sigma}(t,{\bf x})&=&V_{0,\sigma}(t,{\bf x})\,,\label{Vrf}
\end{eqnarray}  
according to Eqs. (\ref{Ur}) and (\ref{Vr}). These are accomplished if we require the normalized modulation functions to satisfy
\begin{equation}\label{rfv}
\lim_{p\to 0} u^{-}_p(t)=\lim_{p\to 0} v^{+}_p(t)=0\,,
\end{equation}
since then the contribution of the matrix $\frac{{p}^i{\sigma}_i}{p}$ is eliminated. We say that these conditions define the {\em rest frame vacuum} which, in general, is different from the adiabatic one as we will see analyzing few examples. 

Concluding we can say that the rest frame vacuum is completely defined by Eqs. (\ref{VU}) and (\ref{uuvv}) and the limits (\ref{Urf}) and (\ref{Vrf}) for the massive Dirac field on any spatially flat FLRW spacetimes.

\section{Examples}

{\em 1. The Minkowski spacetime} is the simplest example of flat manifold where  $t_c=t$ and $a(t)=a(t_c)=1$. The solutions of the systems (\ref{sy1}) and (\ref{sy2}) which satisfy the conditions (\ref{VU}) and  (\ref{uuvv}) read
\begin{eqnarray}
u_p^{\pm}(t)&=&\sqrt{\frac{E(p)\pm m}{2 E(p)}}\,e^{-i E(p) t}\label{UMin}\\
v_p^{\pm}(t)&=&\sqrt{\frac{E(p)\mp m}{2 E(p)}}\,e^{i E(p) t}\label{VMin}
\end{eqnarray}
where $E(p)=\sqrt{p^2 +m^2}$. Thus we recover the standard fundamental spinors of the Dirac theory on Minkowski spacetime \cite{BDR} and we can verify that the conditions (\ref{adi}) and (\ref{rfv}) are satisfied simultaneously. Thus we conclude that in this case  the adiabatic and rest frame vacua {\em coincide}.

{\em 2. The de Sitter expanding universe} is a more delicate example since here the rest frame vacuum of the Dirac field is different from the adiabatic one. This spacetime is defined as the portion of the de Sitter manifold where the scale factor $a(t)=\exp(\omega t)$ depends  on the de Sitter Hubble constant denoted here by $\omega$ \cite{BD}. Consequently, in the conformal chart we have
\begin{equation}
t_c=-\frac{1}{\omega}e^{-\omega t}\in (-\infty, 0]~~~ \to~~~ a(t_c)=-\frac{1}{\omega t_c}\,.
\end{equation}
The normalized solutions in this chart of the system (\ref{sy1c}) or (\ref{sy2c}) can be derived easily obtaining the general normalized solution of the form 
\begin{eqnarray}
u^{+}_p(t_c)&=&\sqrt{-\frac{p t_c}{\pi}}\left[c_1 K_{\nu_{-}}\left(i p t_c\right)+c_2 K_{\nu_{-}}\left(-i p t_c\right)\right]\,,\\
u^{-}_p(t_c)&=&\sqrt{-\frac{p t_c}{\pi}}\left[c_1 K_{\nu_{+}}\left(i p t_c\right)-c_2 K_{\nu_{+}}\left(-i p t_c\right)\right]\,,
\end{eqnarray}
where $K_{\nu_{\pm}}$ are the modified Bessel functions \cite{NIST} of the orders $\nu_{\pm}=\frac{1}{2}\pm i\frac{m}{\omega}$. According to Eq.  ({\ref{H3}) the normalization condition (\ref{uuvv}) is satisfied only if  
\begin{equation}
|c_1|^2+|c_2|^2=1\,.
\end{equation}
The functions $v_p^{\pm}$ result from Eq. (\ref{VU}). The adiabatic vacuum can be defined simply by choosing $c_1=1$ and $c_2=0$ as in Ref. \cite{CD1}. However, this is different from the rest frame vacuum for which we obtain 
\begin{equation}\label{con}
c_1=\frac{e^{\frac{\pi m}{\omega}}p^{-i\frac{m}{\omega}}}{\sqrt{1+e^{\frac{2\pi m}{\omega}}}}\,, \quad c_2=\frac{i\,p^{-i\frac{m}{\omega}}}{\sqrt{1+e^{\frac{2\pi m}{\omega}}}}\,,
\end{equation}
as it results from Eq. (\ref{Urf}) and the behavior (\ref{beh})  of the modified Bessel functions near $z\sim 0$. Hereby,  after using the connection formulas of the modified Bessel functions \cite{NIST}, we obtain the definitive form of the modulation functions of positive frequencies of the rest frame vacuum,
\begin{equation}
u_p^{\pm}(t_c)=\pm \frac{\sqrt{-\pi t_c}\, p^{\nu_-} }{\sqrt{1+e^{\frac{2\pi m}{\omega}}}}\, I_{\mp\nu_{\mp}}(ipt_c)
\end{equation}
which have the  remarkable property  
\begin{equation}
\lim_{t_c\to 0}|u^{+}_p(t_c)|=1\,, \quad \lim_{t_c\to 0}u^{-}_p(t_c)=0\,,
\end{equation}
that may be interpreted as an adiabatic condition at $t\to \infty$ instead of  $t\to- \infty$. The modulation functions of the negative frequencies have to be calculated according to Eq.(\ref{VU}). Note that the constants (\ref{con}) can be seen as the Bogoliubov coefficients of the transformation between the orthonormal bases corresponding of the adiabatic and rest frame vacua. 

{\em 3. A Milne-type spacetime} was studied recently. This is a  spatially flat FLRW spacetime having a Milne-type scale factor $a(t)=\omega  t$ where $\omega$ is a free parameter. Its principal feature is that this is no longer flat as the original Milne's universe, being produced by gravitational sources proportional with $\frac{1}{t^2}$.  On this manifold, it is convenient to use the chart of proper time $\{t,{\bf x}\}$  for $t>0$. In this chart and the diagonal tetrad gauge (\ref{tetrad}), the system (\ref{sy1}) can be analytically solved finding  the general solutions  
\begin{eqnarray}
u_p^{+}(t)&=&\sqrt{\frac{|m| t}{2\pi}}\left\{c_1 \left[K_{\nu_+(p)}(i m t)+ K_{\nu_-(p)}(imt)\right]\right. \nonumber\\
&&~~~~~~~~~\left.+ c_2  \left[K_{\nu_+(p)}(-i m t)- K_{\nu_-(p)}(-imt)\right]\right\}\\
u_p^{-}(t)&=&\sqrt{\frac{|m| t}{2\pi}}\left\{c_1 \left[K_{\nu_+(p)}(i m t)- K_{\nu_-(p)}(imt)\right]\right. \nonumber\\
&&~~~~~~~~~\left.+ c_2  \left[K_{\nu_+(p)}(-i m t)+ K_{\nu_-(p)}(-imt)\right]\right\}
\end{eqnarray}
where now the modified Bessel functions have the orders $\nu_{\pm}(p)=\frac{1}{2}\pm i \frac{p}{\omega}$.  These solutions comply with the normalization condition  (\ref{uuvv})  calculated according to the identity (\ref{H3}) with $\mu=\frac{p}{\omega}$  if 
\begin{equation}
|c_1|^2+|c_2|^2=1\,.
\end{equation}
In the conformal chart $a(t_c)=e^{\omega t_c}$ satisfies the asymptotic condition (\ref{asc}) such that we can introduce the adiabatic vacuum imposing the condition (\ref{adi}) which yields  $c_1=c_2=\frac{1}{\sqrt{2}}$. The rest frame vacuum is given by $c_1=1$ and $c_2=0$ since then  $\lim_{p\to 0} u^-_p(t)=0$. It is worst pointing out that in the momentum-helicity basis and chiral representation of the Dirac matrices (with diagonal $\gamma^5$) the fundamental spinors of the rest frame vacuum take the simple form \cite{prep}
\begin{eqnarray}
U_{\vec{p},\sigma}(x)&=&\sqrt{\frac{mt}{\pi}}\frac{e^{i\vec{p}\cdot\vec{x}}}{[2\pi \omega t]^{\frac{3}{2}}}\left(
\begin{array}{c}
K_{\sigma-i\frac{p}{\omega}}\left(im\,t \right) \xi_{\sigma}(\vec{p})\\
K_{\sigma+i\frac{p}{\omega}}\left(im\,t \right)\xi_{\sigma}(\vec{p})
\end{array}\right)\label{U}\\
V_{\vec{p},\sigma}(x)&=&\sqrt{\frac{mt}{\pi}}\frac{e^{-i\vec{p}\cdot\vec{x}}}{[2\pi \omega t]^{\frac{3}{2}}}\left(
\begin{array}{c}
K_{\sigma-i\frac{p}{\omega}}\left(-im\,t \right) \eta_{\sigma}(\vec{p})\\
-K_{\sigma+i\frac{p}{\omega}}\left(-im\,t \right)\eta_{\sigma}(\vec{p})\\
\end{array}\right)\,,\nonumber\\ \label{V}
\end{eqnarray} 
that can be used in applications. We have thus another example in which the adiabatic and rest frame vacua are different.  

\section{Concluding remarks}

The existence of the new rest frame vacuum is in accordance with the paradigm of the cosmological particle creation where the free fields interact only with the gravity of the background such that different vacua can be selected by the particle detectors.

Another possibility is to consider the quantum theory of interacting fields in which the amplitudes of the quantum transitions have to be calculated using perturbations in terms of free fields \cite{L1,L2,L3,L4,L5,CQED}. The problem is how these free fields may be defined when we may choose from many different vacua.  In   recent applications of the de Sitter QED only the adiabatic vacuum was used so far  \cite{CQED,Cr1,Cr2} but it is possible to consider different vacua for defining the free fields involved in perturbation \cite{L1,L2}. For example, in a collision process we may take the incident beam in the adiabatic vacuum and the target in the rest frame one. However, for the internal lines of the Feynman diagram the rest frame vacuum is the favorite candidate since this can be defined naturally for the massive fields on any spatially flat FLRW spacetime, without requiring supplemental geometric conditions as that of the adiabatic vacuum given by Eq.  (\ref{asc}).

We hope that by using many well-defined vacua we could combine the methods of the perturbative quantum field theory with those of the cosmological particle creation in a more flexible and effective theory of interaction between the quantum matter and gravity.    

\subsection*{Acknowledgments}

This work is partially supported by a grant of  the Romanian Ministry of Research and Innovation, CCCDI-UEFISCDI, project number  PN-III-P1-1.2-PCCDI-2017-0371. 

\appendix

\section{The modified Bessel functions $K_{\nu_{\pm}}(z)$}

According to the general properties of the modified Bessel functions, $I_{\nu}(z)$ and $K_{\nu}(z)=K_{-\nu}(z)$ \cite{NIST}, we
deduce that those used here, $K_{\nu_{\pm}}(z)$, with
$\nu_{\pm}=\frac{1}{2}\pm i \mu$ are related among themselves through
\begin{equation}\label{H1}
[K_{\nu_{\pm}}(z)]^{*}
=K_{\nu_{\mp}}(z^*)\,,\quad \forall z \in{\Bbb C}\,,
\end{equation}
satisfying the equations
\begin{equation}\label{H2}
\left(\frac{d}{dz}+\frac{\nu_{\pm}}{z}\right)K_{\nu_{\pm}}(z)=-K_{\nu_{\mp}}(z)\,,
\end{equation}
and the identities
\begin{equation}\label{H3}
K_{\nu_{\pm}}(i x)K_{\nu_{\mp}}(-i x)+ K_{\nu_{\pm}}(-i x)K_{\nu_{\mp}}(i x)=\frac{\pi}{ |x|}\,,
\end{equation}
that guarantees the correct orthonormalization properties of the fundamental spinors. For 
$|z|\to \infty$  we have  \cite{NIST}
\begin{equation}\label{Km0}
I_{\nu}(z) \to \sqrt{\frac{\pi}{2z}}e^{z}\,, \quad K_{\nu}(z) \to K_{\frac{1}{2}}(z)=\sqrt{\frac{\pi}{2z}}e^{-z}\,,
\end{equation} 
for any $\nu$, while for $z\to 0$ these functions behave as
\begin{equation}\label{beh}
K_{\nu}(z)\sim \frac{1}{2}\Gamma(\nu)\left(\frac{z}{2}\right)^{-\nu}\,.
\end{equation}

\end{document}